\documentclass[twocolumn,prl,aps,showpacs,flushbottom]{revtex4-1}
\usepackage{xspace,amsmath,amsfonts,amsthm,amssymb,amsbsy,graphicx,color,bm}

\newcommand{\ms}[1]{\mbox{\scriptsize #1}}

\definecolor{Red}{rgb}{1,0,0}

\def\dim{\operatorname{dim}}

\def\Span{\operatorname{span}}

\def\Tr{\operatorname{Tr}}

\def\tg{\mathrm{tg}}
\def\su{\mathfrak{su}}

\def\SU{\mathrm{SU}}

\def\A{\mathcal{A}}
\def\B{\mathcal{B}}

\def\D{\mathcal{D}}

\def\F{\mathcal{F}}
\def\G{\mathcal{G}}

\def\M{\mathcal{M}}

\def\P{\mathcal{P}}

\def\T{\mathcal{T}}

\def\J{\mathcal{J}}
\def\S{\mathcal{S}}
\def\K{\mathcal{K}}
\def\ONE{\mathbb{I}}

\begin{document}


\title{Quantum brachistochrone curves as geodesics: \\
obtaining accurate control protocols for time-optimal quantum gates}



\author{Xiaoting Wang$^{1,2}$, Michele Allegra$^{1,3,4}$, Kurt Jacobs$^{2,5}$, Seth Lloyd$^{1,6}$, Cosmo Lupo$^1$, Masoud Mohseni$^7$}  

\affiliation{$^1$Research Laboratory of Electronics, Massachusetts Institute of Technology, Cambridge, MA 02139, USA \\
$^2$Department of Physics, University of Massachusetts at Boston, Boston, MA 02125, USA \\ 
$^3$Dipartimento di Fisica, Universit\`a degli studi di Torino \& INFN, Sezione di Torino, I-10125, Torino, Italy \\ 
$^4$Institute for Scientific Interchange Foundation, I-10126, Torino, Italy \\
$^5$Hearne Institute for Theoretical Physics, Louisiana State University, Baton Rouge, LA 70803, USA \\
$^6$Department of Mechanical Engineering, Massachusetts Institute of Technology, Cambridge, MA 02139, USA \\
$^7$Google Research, Venice, CA 90291, USA $\qquad \qquad   $}

\begin{abstract} 
Most methods of optimal control cannot obtain accurate time-optimal protocols. The quantum brachistochrone equation is an exception, and has the potential to provide accurate time-optimal protocols for essentially any quantum control problem. So far this potential has not been realized, however, due to the inadequacy of conventional numerical methods to solve it. Here, using differential geometry, we reformulate the quantum brachistochrone curves as geodesics on the unitary group. With this identification we are able to obtain a numerical method that efficiently solves the brachistochrone problem. We apply it to two examples demonstrating its power. 
\end{abstract} 

\pacs{03.67.Lx,02.40.-k,02.30.Yy,02.60.Pn} 

\date{August 10, 2014}

\maketitle 

Given a physical quantum device, the study of how to efficiently generate a target unitary gate is important for both fundamental theory and quantum technology.  A powerful approach to this task is to use a time-varying Hamiltonian~\cite{DAlessandro07}. A prescription for varying a Hamiltonian with time to obtain a desired evolution is called a \textit{control protocol}, and a protocol that achieves this task in the minimal time is called \textit{time-optimal}. Since real systems experience noise from their environment, time-optimal protocols often achieve superior fidelities because they minimize the total time of exposure to decoherence. Hence, constructing a time-optimal solution can be considered as a straightforward error-minimization technique for quantum information processing~\cite{Schulte05, Machnes11, Shor95_error, Zanardi97_DFS, Lidar98,Viola99}. We note also that the technique we develop here can be used to improve existing control designs in state-of-the-art experiments~\cite{Blatt11, Zhihui11, Wrachtrup14}. 

From the point of view of numerical optimization methods, finding accurate time-optimal protocols is difficult because it is a two-objective optimization problem: one must maximize the gate fidelity and simultaneously minimize the time taken by the protocol (hereafter the ``protocol time''). 
To find an approximate solution, on the other hand, is relatively easy: one can minimize a weighted sum of the two objectives~\cite{DAlessandro07}, obtaining a sub-optimal protocol, 
or perform multiple optimizations of the gate fidelity, each for a different fixed time, to locate a likely minimal time. But neither method provides solutions of sufficient accuracy that they can be efficiently refined further.  On the other hand, general theories, such as the Pontryagin maximum principle and the geometry of the unitary group, can be used to obtain exact time-optimal solutions, but are applicable only to very specific kinds of problems and constraints~\cite{Margolus98,Khaneja01, sub_riemannian_navin_01,Yuan_05_time,Boscain06, haidong_glaser_cluster_09,Glaser_prl_time_opti_10, Boozer12, Hegerfeldt13,glaser_2013_su2,Billig13}. In view of this, the quantum brachistochrone equation (QBE) was a significant development~\cite{Carlini_branchi,Carlini_unitary}; it could potentially provide time-optimal solutions to any accuracy, and do so under two generally applicable constraints: (i) the system has a finite energy bandwidth; (ii) the Hamiltonian is restricted to a subspace of all Hermitian operators. For any time-optimal problem in the above class, the QBE transforms the optimization problem into that of solving an ODE with boundary values. However the potential of the QBE has not been realized because there exists no numerical method that can solve a high-dimensional boundary-value ODE problem efficiently;
the traditional technique for these problems is called the ``shooting method''~\cite{Press86}, and it usually fails unless seeded with a guess that is sufficiently close to the solution. Even for systems as small as two qubits random guesses are not sufficient, with the result that the QBE has been solved only for a few special examples 
in which analytic solutions are possible~\cite{Carlini_spin1,Carlini_coherence,Carlini_spin2}. 

Here we show that if the norm of the Hamiltonian is bounded (other constraints may also be included, see below) then the problem of finding a control protocol that takes the minimum time can be transformed into a problem of finding a shortest path. If we imagine driving a car over some smooth but undulating terrain (in our case, a differentiable manifold), then if the speed of the car is bounded, and we can always travel at the maximum speed, then we get from A to B fastest by taking the shortest route. We will see that for quantum control it is possible to prove a similar result, and thus connect the brachistochrone problem to a minimum-distance, or geodesic problem. Our second primary result is a numerical method, obtained by exploiting this connection, that can be combined with the shooting method to efficiently solve the brachistochrone equation. This method also suggests a way of identifying globally time-optimal solutions with high confidence, for which the brachistochrone-geodesic connection is essential. 


\textit{Preliminaries~---} To generate a target unitary $U_\tg$ on an $n$-dimensional quantum system, we need to find a time-varying Hamiltonian $H(t)=\sum_m u_m(t) H_m$ such that $U(t)$ satisfies the Schr\"{o}dinger equation
\begin{equation} \label{Eq: schro}
  \dot{U}(t)=-i H(t)U(t)   
\end{equation}
with boundary conditions $U(0)=\ONE$ and $U(T)=U_{\ms{tg}}$  
(we use units such that $\hbar=1$). Here $\{H_m\}$ is the set of Hamiltonian terms that we can physically implement for the system, and $\{u_m(t)\}$ is a set of real functions that will constitute the control protocol. If we neglect a global phase in $U_{\ms{tg}}$, we can restrict $H(t)$ to the $(n^2-1)$-dimensional space of traceless Hermitian matrices, which we will denote by $\M$. We divide $\M$ into $\M=\A\oplus \B$, where $\A=\Span\{A_j\}\equiv\Span\{H_m\}$ is the subspace of Hamiltonians we can implement, 
and $\B=\Span\{B_k\}$ is the subspace we cannot. 
Under the Hilbert-Schmidt product on $\M$, we have $\langle\A,\B\rangle=0$ and $\{A_j,B_k\}$ is an orthonormal basis for $\M$. 
We consider two general physical constraints on $H$. (i) the constraint above that $H(t)=\sum_j\mu_j(t)A_j$ (equivalently $\P_\A(H(t))= H(t)$ or $\P_\B(H(t))= 0$ with $\P_\A$ and $\P_\B$ projectors onto $\A$ and $\B$, respectively); (ii) the energy of the control is bounded via the constraint $||H(t)|| \le E$, where $||\cdot||$ is the Hilbert-Schmidt norm. In addition to (i) and (ii), another natural constraint is that $H(t)$ has a  fixed``free drift'' component that we cannot vary. Here we restrict our analysis to problems without free drift, but similar ideas can de applied to problems that include it. 


\textit{Shortest time vs shortest distance~---} We now show that a bound on the norm of the Hamiltonian is a bound on the speed of evolution. To be precise, a bound on the norm implies that every minimal-time path is a minimal distance path, where distance is defined by the norm. We can show \footnote{Full details are given in the supplemental material.} that for any curve connecting the two fixed points $\ONE$ and $U_\tg$, with $||H(t)|| \le E$, we can merely rescale the Hamiltonian so that the norm is equal to $E$ at all points on the path, and the path is unchanged but takes a shorter time. This means that every minimal-time path has $||H(t)|| = E$. Further, the length of every minimal-time path is given by 
\begin{equation} \label{dist}
  L= \int_0^T \!\!\! ||H(t)|| \, dt = \int_0^T \!\!\! E \, dt = E T , 
\end{equation}
so that minimizing the duration $T$ also minimizes the distance, $L$. Thus we can conclude that the minimum-time curve must also be the minimum-distance curve connecting $\ONE$ and $U_\tg$. Note that this is also true for any particle traveling on a manifold in which only the speed is bounded: the shortest time is achieved by traveling at maximal speed along the shortest path. This is why we can say that a bound on the norm corresponds to a bound on speed.

\textit{Brachistochrone equation~---} 
By virtue of the analysis above, constraint (ii) can be replaced by the equality $\Tr(H^2(t))=E^2$. Now all our constraints are expressed in terms of equalities, and it is possible to derive the minimum-time protocols using variational calculus with Lagrange multipliers. We are now ready to introduce the quantum brachistochrone equation that governs time-optimal protocols under our constraints~\cite{Carlini_unitary}: 
\begin{align}\label{eqn:el_eqn}
\dot H + \sum_k \dot \lambda_k B_k=-i\sum_k \lambda_k[H,B_k] , 
\end{align}
or in component form:
\begin{subequations}
\label{brachi}
\begin{align}
\dot \mu_j  &= i\sum_{k'}\lambda_{k'} \Tr(H[A_j,B_{k'}]), \\   
\dot \lambda_k  &= i\sum_{k'}\lambda_{k'} \Tr(H[B_k,B_{k'}]) . 
\end{align} 
\end{subequations}
The solution to the QBE has the two components $\mu_j(t)$ and $\lambda_k(t)$, where $H(t)=\sum_{\A} \mu_j(t)A_j$ and $\{\lambda_k(t)\}$ are the adjoint Lagrange multipliers, introduced by the Hamiltonian constraints (ii). Together with the Schr\"odinger equation (\ref{Eq: schro}), the QBE (\ref{eqn:el_eqn}) defines a boundary value nonlinear ODE problem with boundary conditions $\ONE$ and $U_\tg$. The time-optimal curve $U(t)$ is uniquely determined by the initial value $\big(\mu_j^0,\lambda_k^0\big):=\big(\mu_j(0),\lambda_k(0)\big)$, and can be written as a flow $U(t)=U(\mu_j^0,\lambda_k^0,t)$. 

To solve the boundary value problem we must first solve the ODEs in (\ref{eqn:el_eqn}) and (\ref{Eq: schro}), parametrizing the solutions as functions of $\big(\mu_j^0,\lambda_k^0\big)$. We then solve the nonlinear equation $U(\mu_j^0,\lambda_k^0,T)=U_\tg$ via a search method for the root $\big(\mu_j^0,\lambda_k^0\big)$. This is the so-called shooting method\cite{Press86}, which works efficiently 
if the initial guess for the root is sufficiently good, but fails if not. As the dimension of the model increases, the complexity of searching for the root 
increases exponentially 
and random guesses become useless, as mentioned above.  

\textit{Geodesic interpretation~---} When the control Hamiltonian $H(t)$ is restricted to the subspace $\A$, the shortest-distance curves that can be generated from $-iH(t)$ are actually geodesics on $\SU(n)$ as a \textit{sub-Riemannian} manifold, rather than a Riemannian manifold. In a sub-Riemannian manifold, distances are measured by allowing only curves tangent to the so-called horizontal subspaces (in this case, $\A$). It is possible to derive the geodesic equation on a sub-Riemannian manifold by using a clever trick: one introduces a Riemannian penalty metric on $\M$~\cite{sub-Rie02} that in an appropriate limit will force $H(t)$ to stay in $\A$. To do this we allow the Hamiltonian to be chosen from the entire space $\M$, so that $H = \sum_{\A} \alpha_j A_j + \sum_{\B} \beta_k B_k$. We now define a new inner product, which we will call the $q$-inner product, by $\langle H_1,H_2 \rangle_q \equiv \sqrt{ \sum_j \alpha_j^{(1)}\alpha^{(2)}_j + q \sum \beta^{(1)}_k \beta^{(2)}_k}$. We also define the $q$-metric for the tangent vectors at $U$ as  $\langle -iH_1U,-iH_2U \rangle_{U,q} \equiv \langle H_1,H_2 \rangle_q$. Accordingly, the length of a curve $U(t)$ under this $q$-metric can be written as $L=\int ||H(t)||_q dt$. This metric applies a penalty proportional to $q$ to the basis operators $B_k$, and thus a penalty to the forbidden subspace. In the limit $q \to  \infty$ we can expect that the geodesics resulting from the $q$-metric (the ``$q$-geodesics'') will be confined exactly to $\A$, and will therefore correspond to the sub-Riemannian geodesics that we seek.  

For a given $q$, the $q$-geodesics are given by the geodesic equation, i.e., the Euler-Lagrange equation for $L$, which is further reduced to the following:
\begin{align}
\label{eqn:geo0}
\G_q(\dot H_q) = i[H_q,\G_q(H)] 
 \end{align}
where $\G_q(H_q)=\P_\A(H_q)+q\P_\B(H_q)$. In the component form, it reads:
\begin{subequations}
\begin{align}
\label{eqn:geo1}
\dot\alpha_j^q&=i\sum_{j'}\alpha_{j'}^q\Tr(H_q[A_j,A_{j'}])+i\sum_{k'}q \beta_{k'}^q\Tr(H_q[A_j,B_{k'}]) \\	\label{eqn:geo2}
q\dot{\beta}_k^q&=i\sum_{j'}\alpha_{j'}^q\Tr(H_q[B_k,A_{j'}])+i\sum_{k'}q\beta_{k'}^q\Tr(H_q[B_k,B_{k'}]) 
\end{align}
\end{subequations}
Alternatively, since the $q$-metric is right-variant, this geodesic equation can be derived from the Euler-Arnold equation~\cite{Arnold66}. This was shown by Nielsen in~\cite{Nielsen06}, where he used the path length in $SU(n)$ to quantify the complexity of quantum computation. 

The geodesic equation (Eq.(\ref{eqn:geo0})) defines a family of $q$-geodesics parametrized by the scalar $q$. The major difference between Eq.(\ref{brachi}) and Eq.(\ref{eqn:geo0}) is that in the latter $H_q(t)=\big( \alpha^q_j(t),\beta^q_k(t) \big)$ is allowed to have components in $\B$, but at a cost determined by $q$. Focusing on the geodesic curves with $||H_q(t)||_q=E$, we expect $\beta_k(t)\to 0$ and $||\P_\A(H)||_{HS}\to ||H_q||_q=E$, as $q\to\infty$. We thus expect to recover the brachistochrone equation from Eq.(\ref{eqn:geo0}) in this limit. Under the assumption 
that the operators $q\beta_k^{q}$ converge as $q\to\infty$, we can prove~\footnote{Full details are given in the supplemental material.} that the first terms in the RHS of Eqs. (\ref{eqn:geo1}) and (\ref{eqn:geo2}) vanish, and in the limit these equations reduce to:
\begin{subequations}
\begin{align}
\dot\alpha^q_j&=i\sum_{k'}q \beta^q_{k'}\Tr(H_q[A_j,B_{k'}])  , \\
q\dot{\beta}^q_k&=i\sum_{k'}q\beta^q_{k'}\Tr(H_q[B_k,B_{k'}]) 
\end{align} 
\end{subequations}
These are identical to the brachistochrone equations (\ref{brachi}) with the replacement $(\alpha_j^q, q\beta_k^{q})$ with $(\mu_j,\lambda_{k} )$. Hence, we have shown that the brachistochrone equation can be considered as the limit of the geodesic equation when $q\to \infty$.
The geometric meaning of the Lagrange multipliers $\{\lambda_k\}$ in Eq.~(\ref{brachi}) is now clear: they are the remaining trails of the vanishing $\beta_k^q$ in $H_q(t)$ along the geodesics. As we now show, the brachistochrone-geodesic connection yields an efficient method to solve the brachistochrone equation by first solving the corresponding geodesic equation. 

\textit{Solving the geodesic equation~---} Together with the Schr\"odinger equation (\ref{Eq: schro})
and the boundary conditions $U(0)=\ONE$ and $U(T)=U_{\ms{tg}}$, the geodesic equation (\ref{eqn:geo0}) defines a boundary-value problem for an ODE, whose solution is fully determined by the initial value $H_q^0\equiv H_q(0)$. 
If we scale $H_q^0$ by a factor $a$ then the total time $T$ scales as $1/a$. For numerical purposes it is convenient to fix $T=1$ to determine $H_q^0$, and afterwards scale the latter by $E/||H_q^0||$ so that its norm is equal to $E$. The total time is then $T = ||H_q^0||/E$. For $q=1$, Eq.~(\ref{eqn:geo0}) reduces to $\dot H_{q=1}(t)=0$ or $H_{q=1}(t) \equiv \bar{H}$. That is, every geodesic on $\SU(n)$ is an evolution under a constant Hamiltonian: $U(t)=e^{-i\bar{H}t}$ with $U_{\ms{tg}}=e^{-i\bar{H}}$. We can find $\bar H$ by taking the logarithm: $\bar H=-i\log(U_{\ms{tg}})$. 
This solution is not unique, however. The solution set is countably infinite: $\{\bar H^{(m)}=-i \log^{(m)}(U_{\ms{tg}}) \}$, $m=1,2,\cdots$, where $\log^{(m)}$ denotes the 
$m^{\mbox{\textit{th}}}$ branch of the logarithm. 

We can now obtain the $q$-geodesics for $q > 1$ by choosing an equally-spaced sequence $\{q_k\}$ with $q_1=1$, $\Delta q=q_{k+1}-q_{k}>0$. As long as $\Delta q$ is sufficiently small, the geodesic solution for $q_k$ is sufficiently close to the $q_{k+1}-geodesic$ that we use it to obtain the latter via the shooting method. Thus, starting from each $H_{q_1}^0=\bar H^{(m)}$, we can find a sequence of geodesic solutions $\{H_{q_k}^0\}$ by consecutively applying the shooting method. Notice that this reasoning holds under the assumption that the geodesic solutions $(H_q(t),U_q(t))$ vary smoothly with respect to $q$. In fact, so long as this is true, there is an even better method for obtaining $H_q^0$ for all $q$ that avoids the shooting method. According to the geodesic deformation technique\cite{Carmo92}, used previously by Dowling and Nielsen\cite{Dowling08b} to study the geodesic equation, when the metric is smoothly varied, the geodesic 
between the two end points is smoothly perturbed in such a way that $H_q^0$ satisfies the differential equation  
\begin{equation}  
\frac{ dH_q^0}{dq}=\mathcal{D}(U_q(t),H_q(t)) , \label{dhdq} 
\end{equation}
where $\mathcal{D}$ is a functional of $U(t),H(t)$ whose form we give in the supplemental material. Coupled with Eqs.(\ref{Eq: schro}) and (\ref{eqn:geo0}), the above differential equation can be solved to obtain $H_q^0$ for $q> q_1$. However, it is possible that a solution of Eq.(\ref{dhdq}) is not defined on the entire domain of $q$, which happens if $\mathcal{D}(U_q(t),H_q(t))$ is undefined at some value $q=q_2 > q_1$. In this case Eq.(\ref{dhdq}) can be used to find $H_q^0$ only up to $q=q_2$. If, on the other hand, the solution of (\ref{dhdq}) can keep extending from $q_1$ to $q\to\infty$, then we can compute $H_q^0$ for any value $q>q_1$ by integrating Eq.(\ref{dhdq}). 
As mentioned above, there are many $H_{q_1}^0$ that can be used to find solutions for $q\to\infty$ ($\beta_k^q(t)\to 0$), and it can be shown from Eq.~(\ref{dhdq}) that $q\beta_k^q(t)$ also converges, and hence the geodesic curve converges to the brachistochrone curve.
In practice, there is a much more efficient method: once we obtain a $q$-geodesic for sufficiently large $q$, this provides a sufficiently good guess for the brachistochrone equation that it can be efficiently solved with the shooting method.



We can now summarize our method for solving the brachistochrone equation: 

\textbf{Step 1}: 
Put all of the solutions of $\bar{H}=-i\log(U_{\ms{tg}})$ into the sequence $\{\bar H^{(m)}\}$, $m=1,2,\cdots$. 

\textbf{Step 2}:  For each $m$: using $H_{q=1}^0=\bar H^{(m)}$ as the initial condition, solve Eq.~(\ref{dhdq}), together with Eqs. (\ref{Eq: schro}) and (\ref{eqn:geo0}), to obtain the family of geodesic solutions $\{ H_q^{(m)}(t),U_q^{(m)}(t)\}$ connecting $\ONE$ and $U_{\ms{tg}}$, parametrized by $q$ and indexed by $m$. 

\textbf{Step 3}: For each $m$: (i) if in Step 2 we are able to solve Eq.~(\ref{dhdq}) up to a value of $q$ for which the $q$-geodesic $U_q^{(m)}(t)$ is sufficiently close to the brachistochrone, then use it as the initial guess to solve the brachistochrone equation (Eq.(\ref{eqn:el_eqn})). (ii) If the solution to Eq.~(\ref{dhdq}) terminates before a sufficiently large value of $q$ can be obtained (that is, $H_q^{(m)}$ cannot be calculated for $q\to\infty$),
then abandon that solution and move to the next value of $m$. 

\textbf{A special case:} When $[\P_\A(H_{q=1}^0),\P_\B(H_{q=1}^0)]=0$ the derivative $dH_q^0/dq|_{q=1}$ vanishes, and we cannot obtain $H_{q>1}^0$ from $H_{1}^0$ using Eq.(\ref{dhdq}). Nevertheless, for $q\ne 1$ but near to 1, the shooting method with a random initial guess is still effective at solving for $H_{q}(t)$ in Eq.(\ref{eqn:geo0}). From there we can obtain the geodesics for larger values of $q$ by integrating Eq.(\ref{dhdq}).  


At first sight, the above method may seem inefficient because there are an infinite number of geodesic families, one for each $m$. In practice we do not need to calculate the $q$-geodesics for every $m$ to find one or more globally time-optimal protocols. Simulation results suggest that i) within each geodesic family $m$ that can extend to $q\to\infty$ the protocol time is monotonically increasing with $q$, and ii) the ordering of the protocol times with $m$ is independent of $q$. Thus, if we have a rough estimate $T^\ast$ of the minimum protocol time, it is sufficient to consider the initial solutions $\bar H^{(m)}$ with $ T^{(m)} < T^\ast $. As discussed in the introduction, there are simple methods that can be used to find rough estimates of the global minimum time, and thus provide a $T^*$. In principle, the Hamilton-Jacobi-Bellman equation~\cite{Kirk04} could also be used to determine which brachistochrone solution, obtained numerically, corresponds to the global minimum time. 


\textit{Examples~---} To demonstrate our method, we consider the following two-qubit model 
\begin{equation}
H = \hbar \! \sum_{l,m}  \! \omega_m^{(l)} \sigma_m^{(l)} + \hbar \kappa \sum_{m} \sigma_m^{(1)}\otimes\sigma_m^{(2)} , 
\end{equation}
where $\sigma_{m}^{(l)}$, $m = x, y, z$, $l=1,2$ are the Pauli operators for the $l^{\ms{\textit{th}}}$ qubit. We assume that we can vary the six parameters $\{\omega_m^{(l)}\}$ and the inter-qubit coupling rate $\kappa$. The accessible and forbidden spaces for this model are thus $\A=\Span\{\sigma_m^{(l)},\sigma_m^{(1)}\otimes\sigma_m^{(2)}\}$ and $\B=\M/\A$, respectively. 

\begin{figure}[t]
\leavevmode\includegraphics[width=0.9\hsize]{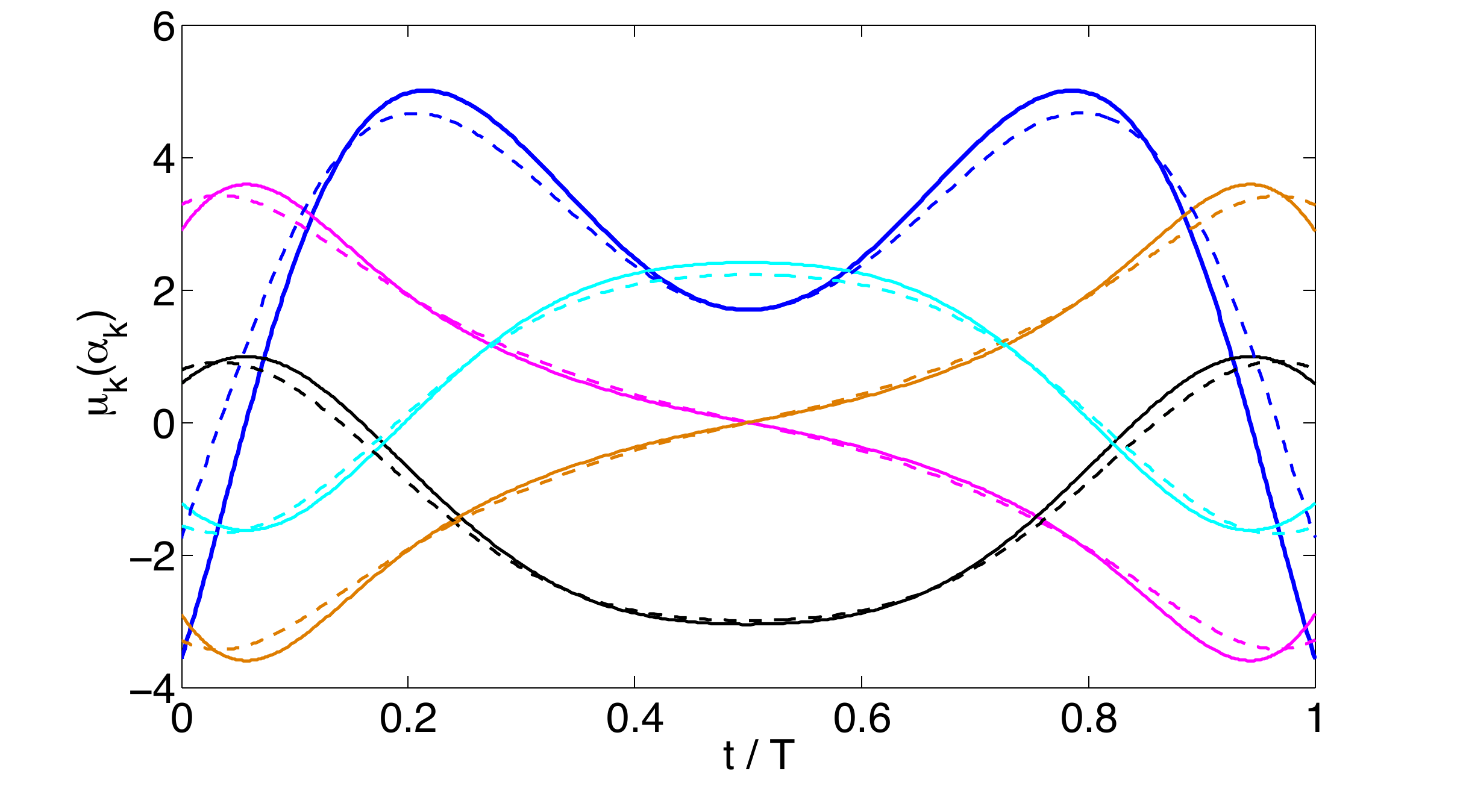}
\caption{Here we show the seven control functions $\mu_k(t)$, $k=1,\cdots,7$, that implement the minimal-time CNOT gate (solid curves) for a given 2-qubit interaction, along with the seven functions $\alpha_k(t)$ for the approximate protocol which is a crucial step in obtaining it (dashed curves). Two pairs of control functions are identical, so only five distinct curves appear in the plot. The approximate protocol is a geodesic for a metric with $q=100$ (see text).} 
\label{fig1} 
\end{figure} 

As our first example, we choose the target unitary, $U_\tg$, to be a randomly selected 2-qubit operator in $\SU(4)$. We give the explicit expression for  $U_\tg$ in the supplemental material. We first attempt to use the shooting method to directly solve Eq.(\ref{brachi}). We have tried one hundred different randomly chosen initial guesses, and find that every try fails.
We then apply the new method presented above.
We initially fix $T=1$. For $q=1$, solving $H_{q=1}^0 = -i\log(U_{\ms{tg}})$ gives a sequence of solutions $\{\bar H^{(m)}\}$ with norms in a nondecreasing order. 
For each $\bar H^{(m)}$, we integrate Eq.(\ref{dhdq}) to get the geodesic solution for sufficiently large $q$. For $\bar H^{(1)}$ and $\bar H^{(2)}$, we find Eq.(\ref{dhdq}) cannot be integrated for sufficiently high $q$, and hence they have to be abandoned. For $H_{q=1}^0=\bar H^{(3)}$, Eq.(\ref{dhdq}) can be integrated from $q=1$ to $q=100$, with geodesic solution denoted as $H_q(t)$, satisfying $||\P_\B(H_{q=1})||/||H_{q=1}||=0.42$, and $||\P_\B(H_{q=100})||/||H_{q=100}||=0.03$. Denote the components of $H_q(t)$ as $(\alpha_j^{q}(t),\beta_k^{q}(t))$. 
Due to the brachistochrone-geodesic connection, the geodesic curve $H_{q=100}(t)$ provides a good approximated solution for Eq.(\ref{brachi}): $(\mu_j,\lambda_k)\equiv (\alpha_j^{q},q\beta_k^{q})$, with fidelity $0.9916$, indicating that it is close to a true brachistochrone. We then use this approximated solution to seed the shooting method to solve Eq.(\ref{brachi}). This attempt succeeds, resulting in a minimum-time protocol $H(t)=\P_\A(H(t))$ with infidelity $\varepsilon \equiv 1 - F  < 10^{-10}$ and $||H||=6.69$, so that after rescaling we obtain   
a protocol time of $T=6.69/E$. On the other hand, a weighted-sum optimization gives an upper bound $T^\ast=6.8/E$.
Repeating the above procedure for larger values of $m$, with $||\bar H^{(m)}||<6.8$, we can numerically find other brachistochrone solutions. All those that we have calculated give $T^{(m)}>6.69/E$. As regards the run-time, solving Eq.(\ref{dhdq}) from $q=1$ to $q=100$ took approximately 130 minutes on our machine, and solving Eq.(\ref{brachi}) via the shooting method using $H_{q=100}$ took about 2 minutes.

As our second example we find a time-optimal implementation of the CNOT gate~\cite{mikeandike}. To do so we first add a global phase of $\pi/4$ to the standard CNOT, so that $U_{\ms{tg}}=e^{i \pi/4}U_{\ms{CNOT}}$ is in $\SU(4)$.  Since $[\P_\A(H_{q=1}^0),\P_\B(H_{q=1}^0)]=0$, this problem is an example of the special case noted above. We must therefore solve the geodesic equation for a value of $q'>1$ using the shooting method. Here, at $q' = 5$, we solve Eq.(\ref{eqn:geo0}) and obtain the geodesic solution $H_{q=5}(t)$. We then integrate Eq.(\ref{dhdq}) from $q=5$ to $q=100$, obtaining $H_{q=100}(t)$ with a fidelity of $0.9978$. The third and final step then gives us a brachistochrone solution $H(t)$ with infidelity $\varepsilon < 10^{-10}$ and protocol time $T = 5.75/E$. We plot the 7 components $(\mu_j(t))$ of the brachistochrone $H(t)$ in Fig.~\ref{fig1}, along with the seven components $(\alpha_j(t))$ of the geodesic solution $H_{q=100}(t)$ for comparison. We see that when $q$ is sufficiently large, the geodesic solution is close to the brachistochrone solution. 


\textit{Acknowledgements:} XW and KJ were partially supported by NSF project PHY-1005571 and the ARO MURI grant W911NF-11-1-0268. KJ was also partially supported by NSF project PHY-1212413. 

\bibliography{report}

\widetext
\clearpage
\begin{center}
\textbf{\large Supplement of ``Quantum brachistochrone curves as geodesics: \\
obtaining accurate control protocols for time-optimal quantum gates"}\\
\vspace{0.5cm}
{Xiaoting Wang, Michele Allegra, Kurt Jacobs, Seth Lloyd, Cosmo Lupo, Masoud Mohseni} \\
{\small (Dated: August 10, 2014)}
\end{center}
\vspace{0.01cm}
{\small In this supplemental material, we present a detailed proof of the brachistochrone-geodesic connection, and give two examples based on a two-qubit model demonstrating how to apply the proposed algorithm to find accurate time-optimal control protocols. }

\subsection{Shortest time vs shortest distance}

Let $(H(t),U(t))$ be a control protocol satisfying constraints (i) and (ii), such that $\P_\A(H(t))=H(t)$, $||H(t)||\le E$ and $U(t)=\T\big(e^{-i\int_0^{t}H(t)dt}\big)$ with $U(0)=\ONE$, $U(T)=U_\tg$, and $\T$ as the time-ordering operator. Define a new time parameter $s$, which is a monotonic function of the old time variable $t$:
\begin{align*}
s(t)\equiv \frac{1}{E}\int^t_0 ||H(t)||dt, \text{ or } ds=\frac{1}{E}||H(t)||dt
\end{align*}
Since $s$ is a monotonic function of $t$, $H(t)$ can be expressed a function of $s$: $H(s)=H(t(s))$. We can define new Hamiltonian $\bar H$ as a function of $s$: $\bar H(s)\equiv \frac{E H(t(s))}{||H(t(s))||}$ satisfying $||\bar H||=E$ and
\begin{align*}
\T\big(e^{-i\int \bar H(s)ds}\big)=\T\big(e^{-i\int \frac{E H(t)}{||H(t)||}\frac{1}{E}||H(t)||dt}\big)=\T\big(e^{-i\int H(t)dt}\big)
\end{align*}
with $U(s=0)=U_0$ and $U(s=s(T))=U_{\ms{tg}}$. Notice that $\bar H$ also satisfies constraints (i) and (ii).  Moreover, 
\begin{align*}
s(T)=\frac{1}{E}\int_0^{T} ||H(t)||dt \leq \frac{\max_t||H(t)||}{E} \int_0^{T} dt \leq T 
\end{align*}
and $s(T)<T$ if $||H(t)||<E$ during some interval in $[0,T]$. Hence, a time-optimal solution $(H(t),U(t))$ must satisfy $||H(t)||\equiv E$. 

On the other hand, under the right-invariant norm induced by the Hilbert-Schmidt norm, the length of the curve $U(t)$ is
\begin{equation}
L = \int_0^T \!\!\! ||H(t)|| \, dt = \int_0^T \!\!\! E \, dt = E T .  
\end{equation}
Therefore, the shortest-time solution must also be the shortest-distance solution connecting $\ONE$ and $U_\tg$. 

\subsection{Euler-Lagrange equation on $\SU(N)$}

Next we show that both the brachistochrone equation and the geodesic equation can be
obtained as Euler-Lagrange equations on the differential manifold $\SU(N)$, through the variation of some action $\S=\int_\gamma \mathcal{L} dt$. 

The Euler-Lagrange equation is essential in variational calculus. In general, given a manifold $M$, the Lagrangian $\mathcal{L}$ is a function defined on the tangent bundle $TM$. Denoting the local coordinates near a point $p\in M$ as $X^m(p)$, and the basis of the tangent space at $p$ as $\partial_m^{(p)}\equiv \frac{\partial}{\partial X^m}|_p$, $\mathcal{L}$ at $p$ can be expressed as a function of $(X^m(p),X^m(p))$. Accordingly, along a curve $\gamma(t)=X^m(t)$, $\mathcal{L}$ can be written as a function of $(X^m(t), \dot{X}^m(t))$. The action functional $S[\gamma]=\int_{\gamma}dt\ \mathcal{L}(t)$
is minimized along the curve $\gamma$ such that the  Euler-Lagrange equations are satisfied:
\begin{align*}
\frac{d}{dt}\Big(\frac{\partial\mathcal{L}}{\partial\dot{X}^{m}}\Big)=\frac{\partial\mathcal{L}}{\partial X^{m}}
\end{align*}
In the following,  we will consider $M=\SU(N)$. We consider a special type of Lagrangian that only depends on $H(t)=-i\dot{U}(t)U(t)$ at every point of a curve $U=U(t)$.
This peculiar dependence ensures that 
the action $\S$ is invariant under the right-translation of the curve, which considerably simplifies the equation of motion. \\
A local region near any point $U$ on $\SU(N)$ can be parametrized by some coordinates $X^m$, and the curve $U(t)$ within that region can be represented by $X^m(t)$.
Specifically, we parametrize $\SU(N)$ as $U=e^{-iX}$ where $X$ is a Hermitian matrix $X = \sum_m X^m C_m$, where $\{ -iC_m \} = \{-iA_j, -iB_k\} $ is a basis of 
$\su(n)$, orthonormal with respect to the Hilbert-Schmidt scalar product.
For a curve $U(t)=e^{-iX(t)}$ we have:
\begin{align*}
U(t+\Delta t)=e^{-iX(t+\Delta t)}=e^{-i(X(t)+\dot{X}(t)\Delta t)}+\mathcal{O}(\Delta t^{2})
\end{align*}
This formula can be reexpressed upon introducing $H(t)=i\dot{U}(t)U^\dagger(t)$,  
\begin{align*}
U(t+\Delta t)=e^{-iH(t)\Delta t}e^{-iX(t)}+\mathcal{O}(\Delta t^{2})
\end{align*}
Under right translations, $U=e^{-iX}\mapsto UV=e^{-iX'} $, so  $X$ and $\dot{X}$ are changed 
as $X \mapsto X' $ , $\dot{X} \mapsto \dot{X}'=\frac{dX'}{dX} \dot{X}$, but $H$ remains invariant, $\dot{U}V(UV)^\dagger=\dot{U}V$. In fact, $H=\sum_m H^m \sigma_m$ can
be regarded as a right-invariant representation of the tangent vector.\par
The relation between $H,X,\dot{X}$ can be expressed as $H=\phi(X,\dot{X})$
and can be computed as follows. Up to corrections of order $\mathcal{O}(\Delta t^{2})$ we have:
\begin{align*}
\log(e^{-iH\Delta t}e^{-iX})=-i(X+\dot{X}\Delta t)+\mathcal{O}(\Delta t^{2})
\end{align*}
Upon using the logarithmic form of the BCH formula, 
\begin{align*}
\log(e^{A}e^{B})=A+\frac{ad_{A}e^{ad_{A}}}{e^{ad_{A}}-1}B+\mathcal{O}(B^{2}) \rightarrow \log(e^{-iH\Delta t}e^{-iX})=-\log(e^{iX}e^{iH\Delta t})=-iX-\frac{iad_{X}e^{iad_{X}}}{e^{iad_{X}}-1}iH\Delta t+\mathcal{O}(\Delta t^{2})
\end{align*}
where $ad_A=[A,\cdot]$ and after simple algebra we can get:
\begin{align*}
H=\phi(X,\dot{X})=(ad_{X})^{-1}i(e^{-iad_{X}}-1)\dot{X}
\end{align*}

Suppose now that the Lagrangian  depends on $X$ and $\dot{X}$ only though the combination $\phi(X,\dot{X})$.
Then the Lagrangian is invariant under right translations. As a particular 
case one may have a right-invariant metric, $\mathcal{L}(X,\dot{X})=g(H,H)=g(\phi(X,\dot{X}),\phi(X,\dot{X}))$ where 
$g$ is a metric on $\su(n)$. The right-invariance of the Lagrangian has important consequences
as for the equations of motion. We have indeed:
\begin{align*}
\frac{\partial\mathcal{L}}{\partial X^{m}}|_{X_0}=\frac{\partial\mathcal{L}}{\partial{H}^{r}}\frac{\partial H^r}{\partial{X}^{m}}|_{X_0},\quad
\frac{\partial\mathcal{L}}{\partial\dot{X}^{m}}|_{X_0}=\frac{\partial\mathcal{L}}{\partial{H}^{r}}\frac{\partial H^r}{\partial{\dot{X}}^{m}}|_{X_0}
\end{align*}
Now, because of the right invariance of $H$ we have
\begin{align*}
\partial{H}^r/\partial{X}^m|_{X_0}=\partial{H}^r/\partial{X'}^m|_{X'_0}, \quad \partial{H}^r/\partial{\dot{X}^m}|_{X_0}=\partial{H}^r/\partial{\dot{X}'^m}|_{X'_0}
\end{align*}

Upon translating by $e^{iX_0(t)}$ we can thus evaluate $\partial{H}/\partial{X},\partial{H}/\partial{\dot{X}}$
at the intentity $\mathbb{I}$ corresponding to $X'_0=0$. For $X'$ in the vicinity of the origin, we have:
\begin{align*}
\frac{ad_{X'}e^{iad_{X'}}}{e^{iad_{X'}}-1} = -i(1+\frac{i}{2}ad_{X'})+\mathcal{O}(X'^{2})
\end{align*}
whence we can derive the relations between $H$ and $X'$:
\begin{align}
\dot{X'}=H+\frac{i}{2}[X',H]+\mathcal{O}(X'^{2}) \\
H=\dot{X'}-\frac{i}{2}[X',\dot{X'}]+\mathcal{O}(X'^{2})
\end{align}
Upon expanding the latter equation in the Pauli basis,
\begin{align*}
H^{r}=\dot{X'}^{r}-\frac{i}{2}[X',\dot{X}']^{r}+\mathcal{O}(X'^{2})=\dot{X}'^{r}-\frac{i}{2}c_{mnr}X'^{m}\dot{X}'^{n}+\mathcal{O}(X'^{2})
\end{align*}
where $c_{mnr}$ are the structure constants of the group
(for $\SU(N)$ they are completely antisymmetric). The terms in the Euler-Lagrange equations become:
\begin{align*}
\frac{\partial\mathcal{L}}{\partial\dot{X}^{m}} & =  \frac{\partial\mathcal{L}}{\partial H^{r}}
\frac{\partial H^{r}}{\partial\dot{X}^{m}}=  \frac{\partial\mathcal{L}}{\partial H^{r}}
\frac{\partial H^{r}}{\partial\dot{X}'^{m}}= \frac{\partial\mathcal{L}}{\partial H^{m}}-\frac{i}{2}\frac{\partial\mathcal{L}}{\partial H^{r}}
c_{nmr}X'^{n}+\mathcal{O}(X'^{2})\\
\frac{d}{dt}\Big(\frac{\partial\mathcal{L}}{\partial\dot{X}^{m}}\Big) & =
\frac{d}{dt}\Big(\frac{\partial\mathcal{L}}{\partial H^{m}}\Big)-\frac{i}{2}\frac{d}{dt}
\Big(\frac{\partial\mathcal{L}}{\partial H^{m}}\Big)c_{nmr}X'^{nn} -  \frac{i}{2}\frac{\partial\mathcal{L}}{\partial H^{r}}c_{nmr}\dot{X}'^{n}+\mathcal{O}(X') \\
\frac{\partial\mathcal{L}}{\partial X^{m}} & =\frac{\partial\mathcal{L}}{\partial H^{r}}\frac{\partial H^{r}}{\partial X^{m}} =\frac{\partial\mathcal{L}}{\partial H^{r}}\frac{\partial H^{r}}{\partial X'^{m}}=
\frac{\partial\mathcal{L}}{\partial H^{r}}\big(-\frac{i}{2}c_{mnr}\dot{X}'^{n}+\mathcal{O}(X')\big)
\end{align*}
Hence, the Euler-Lagrange equations at $X(t)$ become (taking into account $X'=0, \dot{X'}|_{X'=0}=H$):
\begin{align*}
\frac{d}{dt}\Big(\frac{\partial\mathcal{L}}{\partial H^{m}}\Big)-\frac{i}{2}\frac{\partial\mathcal{L}}{\partial H^{r}}c_{nmr}H^{n}+\frac{i}{2}\frac{\partial\mathcal{L}}{\partial H^{r}}c_{mnr}H^{n}=0
\end{align*}
Upon introducing $F^{m} \equiv \frac{\partial\mathcal{L}}{\partial H^{m}}$, and using $c_{mnr}=-c_{nmr}$, we get:
\begin{equation}
\dot{F}^{m}-iF^{r}c_{nmr}H^{n}=0
\end{equation}
or, using $F=\sum_{m}F^{m}C^{m}$, $H=\sum_{m}H^{m}C^{m}$,
and $F^{r}c_{nmr}H^{n}=F^{r}c_{rnm}H^{n}=[F,H]^{m}$, we have
\begin{equation}\label{el}
\dot{F}=-i[H,F]
\end{equation}

As the first example, let's derive the brachistochrone equation under constraints (i) and (ii). The admissible Hamiltonian $H(t)$ is within the subspace $\A$ satisfying $||H||=E$, with $\M=\A\oplus \B$. The action functional for total time under the constraints can be chosen as: $S=\int_{0}^T\mathcal{L}\, dt$, with
\begin{equation*}
\mathcal{L}=1+\frac{1}{2}\lambda_0\big(\Tr(H^2)-E^2\big)+ \sum_k \lambda_k \Tr(HB_k)
\end{equation*}
where $\{\lambda_0,\lambda_j\}$ are Lagrange multipliers, and $\{B_k\}$ is the basis of $\B$. From $F\equiv \frac{\partial \mathcal{L}}{\partial H}=\lambda_0 H+\sum_k\lambda_k B_k$, Eq.~(\ref{el}) gives:
\begin{equation*}
\dot \lambda H+\lambda \dot{H}+\sum_k\dot{\lambda}_k B_k = -i[H,\sum_k\lambda_k B_k]
\end{equation*}
Multiplying by $H$ and taking the trace of the two sides, we find $\dot \lambda E^2+\lambda \Tr(\dot{H}H)=0$. From $\Tr(H^2)=E^2$, we have $\Tr(\dot H H)=0$, and hence we have $\dot \lambda=0$ or $\lambda$ being a constant. We can rescale $\lambda_j$ such that $\lambda_0$ can be chosen to be 1, and then the above equation becoms:
\begin{equation}\label{brachi0}
\dot{H}+\sum_k\dot{\lambda}_k B_k = -i[H,\sum_k\lambda_k B_k]
\end{equation}
which is the quantum brachistochrone equation originally derived in \cite{Carlini_unitary}. 

As the second example, we discuss the geodesic equation for the $q$-metric as defined in the main paper. Under the $q$-metric, the length of a curve $U(t)$ can be written as: 
\begin{align*}
\int_0^T\langle H(t), H(t)\rangle_q \, dt
\end{align*}
For a rescaled Lagrangian $\mathcal{L}=\frac{1}{2}\langle H, H\rangle_q=\frac{1}{2}\Tr(H\G_q(H))$ where $\G_q(H)\equiv\P_\A(H)+q\P_\B(H)$, we have $F\equiv \frac{\partial \mathcal{L}}{\partial H}=\G_q(H)=\sum_{\A} \alpha_j A_j + q\sum_{\B} \beta_k B_k$. Then Eq.(\ref{el}) becomes: 
\begin{align}
\label{eqn:geo1_SM}
\G_q(\dot H_q) = i[H_q,\G_q(H)] 
 \end{align}
where the index $q$ indicates that the extremal solution $(U_q(t),H_q(t))$ is under the given $q$-metric. This is the geodesic equation, which is equivalent to the Euler-Arnold equation originally derived in~\cite{Nielsen06}. For different values of $q$, the geodesic solutions are different, but we can smoothly vary a geodesic curve over different values of $q$. This is commonly known as geodesic variation or deformation technique, and will be discussed in the following. 

\subsection{Brachistochrone-geodesic connection}

Under constraints (i) and (ii), we assume $H(t)$ has a decomposition $H(t)=\sum_j \mu_j(t)A_j$, where $\{A_j\}$ forms a basis for the physical accessible subspace $\A$ and $\{B_j\}$ is the basis for the forbidden subspace $\B$. 

For the brachistochrone equation (\ref{brachi0}), multiplying by $A_j$ and $B_k$ respectively on both sides and tracing, we get the component form of the brachistochrone equation:
\begin{subequations} \label{eqn:brachicomp}
\begin{align}
\dot \mu_j  &= i\sum_{k'}\lambda_{k'} \Tr(H[A_j,B_{k'}])\\   
\dot \lambda_k  &= i\sum_{k'}\lambda_{k'} \Tr(H[B_k,B_{k'}])
\end{align}
\end{subequations}
Under the $q$-metric, let the geodesic solution be $H_q(t)=\sum_j \alpha_j^q(t)A_j+\sum_k \beta_j^q(t)B_k$, Analogously, multiplying by $A_j$ and $B_k$ respectively on both sides of Eq.(\ref{eqn:geo0}) and tracing, 
we get the component form of the geodesic equation:  
\begin{align} \label{eqn:equiv_1}
\dot{\alpha}_j^q &= i \sum_{j'}\alpha_{j'}^q\Tr(H_q[A_j,A_{j'}])+i \sum_{k'}q\beta_{k'}^q\Tr(H_q[A_j,B_{k'}]) \\  \label{eqn:equiv_2}
q\dot{\beta}_k^q & =i \sum_{j'}\alpha_{j'}^q\Tr(H_q[B_k,A_{j'}])+i \sum_{k'}q\beta_{k'}^q\Tr(H_q[B_k,B_{k'}])
\end{align}
For different values of $q$, there are different geodesic solutions $(\alpha^q_j,\beta^q_k)$ that connect $\ONE$ and $U_{\ms{tg}}$. 
When $q\to\infty$, in order to have well-defined limiting solutions, $(\alpha^q_j,\beta^q_k)$ must converge to finite values.
Actually, for $(\alpha^q_j)$ to converge, a stronger convergence condition on $(\beta^q_j)$ is required. Define \[
\Lambda^q_{k}(t)=q\beta^q_{k}(t)\]
As $q\to \infty$, $\Lambda^q_{k}(t)$ either diverges or converges to a finite value. 
If $\Lambda^q_{k}(t)$ diverges, according to the above equation, $\dot{\alpha}_{j}^q$ also diverges, so that $\alpha^q_j$ has no well-defined limit.
Hence $(\alpha^q_j)$ coverges only if $\Lambda^q_{k}(t)$ converges as $q\to \infty$, which is equivalent to $\beta^q_k=\mathcal{O}(\frac{1}{q})$. 
To sum up, we make the following stronger assumption:\\
 (\textbf{A1}) $(\alpha^q_j,\Lambda^q_k)\equiv (\alpha^q_j,q\beta^q_k)$ 
converge as $q \to \infty$. 

Consider the first term in (\ref{eqn:equiv_1}). It can be rewritten as
\begin{equation*}
i\sum_{j'l}\alpha_{j'}^q\alpha_l^q \Tr\big(A_l[A_j,A_{j'}]\big) + i\sum_{j'm}\alpha_{j'}^q\beta_m^q \Tr\big(B_m[A_j,A_{j'}]\big),	
\end{equation*}
whose first term can be rewritten as
\begin{equation*}
i\sum_{j'l}\alpha_{j'}^q\alpha_l^q \Tr\big(A_l[A_j,A_{j'}]\big) = i\Tr\big(\sum_{j'l}\alpha_{j'}^q\alpha_l^q[A_l,A_{j'}]A_j\big)
\end{equation*}
By antisymmetry, $\sum_{j'l}\alpha_{j'}^q\alpha_l^q[A_l,A_{j'}]$ vanishes and we are left with
\begin{equation*}
i\sum_{j'm}\alpha_{j'}^q\beta_m^q \Tr(B_m[A_j,A_{j'}])	
\end{equation*}
In the limit,  $q \to \infty$, under the assumption that $\beta_m^q = \mathcal{O}(1/q)$ , this term converges to $0$.  Hence, under (A1), in the limit $q\to\infty$ the first term of (\ref{eqn:equiv_1}) converges to zero and (\ref{eqn:equiv_1}) is reduced to:
\begin{equation*}
\dot{\alpha}_j^q = i \sum_{k'}\Lambda_{k'}^q\Tr(H_q[A_j,B_{k'}]) 
\end{equation*}
Next, consider the first term in equation (\ref{eqn:equiv_2}). It can be rewritten as
\begin{equation*}
i \sum_{j'l}\alpha_{j'}^q\alpha_l^q\Tr(A_l[B_k,A_{j'}])+ i \sum_{j'm}\alpha_{j'}^q\beta_m^q\Tr(B_m[B_k,A_{j'}])
\end{equation*}
The first term can be rewritten as
\begin{equation*}
i\sum_{j'l}\alpha_{j'}^q\alpha_l^q \Tr(A_l[B_k,A_{j'}]) = i\Tr(\sum_{j'l}\alpha_{j'}^q\alpha_l^q[A_l,A_{j'}]B_k)
\end{equation*}
By antisymmetry, $\sum_{j'l}\alpha_{j'}^q\alpha_l^q[A_l,A_{j'}]$ vanishes and we are left with
\begin{equation*}
i \sum_{j'm}\alpha_{j'}^q\beta_m^q\Tr(B_m[B_k,A_{j'}])
\end{equation*}
In the limit,  $q \to \infty$, under the assumption that $\beta_m^q = \mathcal{O}(1/q)$ , this term converges to $0$.  Hence, under (A1), in the limit $q\to\infty$ the first term of (\ref{eqn:equiv_2}) converges to zero and (\ref{eqn:equiv_2}) is reduced to:
\begin{equation}
\dot{\Lambda}_k^q  = i \sum_{k'}\Lambda_{k'}^q\Tr(H_q[B_k,B_{k'}])
\end{equation}
%
%
To sum up, as $q\to\infty$, under the assumption $\beta^q_k=\mathcal{O}(\frac{1}{q})$ for any $B_k \in \mathcal{B}$, the geodesic equation converges to the following equation: 
\begin{subequations}
\label{eqn:nielsen-limit}
\begin{align}
\dot{\alpha}_j^q &= i \sum_{k'}\Lambda_{k'}^q\Tr(H_q[A_j,B_{k'}]) \\  
\dot{\Lambda}_k^q & = i \sum_{k'}\Lambda_{k'}^q\Tr(H_q[B_k,B_{k'}])
\end{align} 
\end{subequations}
which have the same form of the brachistochrone equations (\ref{eqn:brachicomp}) upon replacing $(\alpha_j^q,\Lambda_k^q)$ with $(\mu_j,\lambda_k)$.
Thus, we have shown that the quantum brachistochrone equation can be considered as the limit of a bundle of geodesic equations, parametrized by $q$, under the assumption that $\beta^q_k=\mathcal{O}(\frac{1}{q})$ and $\alpha^q_j$ converges as $q\to\infty$. Moreover, if $(\alpha_j^q, \beta_k^q)$ is the corresponding solution of the geodesic equation for a sufficiently large $q$, then $(\alpha_j^q, \Lambda_k^q)=(\alpha_j^q, q\beta_k^q)$ approximates the solution $(\mu_j,\lambda_k)$ for the corresponding limiting brachistochrone equation. 


\subsection{Geodesic deformation technique}

The following derivation was originally given in~\cite{Dowling08b}. For a given value $q$, let $U_q(t)$ and $H_q(t)$ satisfy the geodesic equation that connects $\ONE$ and $U_{\ms{tg}}$. We aim to express the geodesic $U_{q+dq}(t)$ and $H_{q+dq}(t)$ for the value $q+dq$, based on $U_q(t)$ and $H_q(t)$. Let us assume there is a matrix function $J(t)$ such that the relation between $U_q(t)$ and $U_{q+dq}(t)$ can be written as:
\begin{align*}
 U_{q+dq}(t)=U_{q}(t)e^{-iJ(t)dq}
\end{align*}
where $J(0)=J(T)=0$ so that $U_{q+dq}(t)$ also connects $\ONE$ and $U_{\ms{tg}}$. From this relation and the Schrodinger equation, 
\begin{align*}
 \dot U_{q}(t)&=-iH_{q}(t)U_{q}(t)\\
 \dot U_{q+dq}(t)&=-iH_{q+dq}(t)U_{q+dq}(t),
\end{align*}
we have
\begin{align*}
 \dot U_{q+dq}(t)=\dot U_{q}(t)e^{-iJ(t)dq}+U_{q}(t)\big(-idq\dot J(t)+\mathcal{O}(dq^2)\big),
 \end{align*}
which gives 
\begin{align*} 
-iH_{q+dq}(t)U_{q+dq}(t)= -iH_{q}(t)U_{q+dq}(t)+U_{q}(t)\big(-idq\dot J(t)+\mathcal{O}(dq^2)\big).
\end{align*}
Hence, we have
\begin{align*}
 &-iH_{q+dq}(t)= -iH_{q}(t)+U_{q}(t)\big(-idq\dot J(t)\big)U^\dag_{q+dq}(t)+\mathcal{O}(dq^2)
 = -iH_{q}(t)+U_{q}\big(-idq\dot J(t)\big)U^\dag_{q}+\mathcal{O}(dq^2)
\end{align*}
and taking the limit $dq\to 0$ we obtain
\begin{align} \label{Eq: dhdq}
 \frac{ dH_{q}(t)}{dq}=U_{q}(t)\dot J(t)U^\dag_{q}(t)\equiv K(t)
 \end{align}
where for simplicity we have defined a new operator $K(t)=U_{q}(t)\dot J(t)U^\dag_{q}(t)$. In particular, we have $\dot J(0)=K(0)=\frac{ dH_{q}(0)}{dq}$. Notice that from the definition, both $J(t)$ and $K(t)$ are also functions of $q$, but for simplicity we will omit these lower indices in the following. 
Moreover, defining a superoperator $\F_q=\P_\A+q^{-1}\P_\B$, i.e., $\F_q=\G_q^{-1}$, we can rewrite the geodesic equation~(\ref{eqn:geo1_SM}) as 
\begin{align*}
\dot H_q=-i\F_q([H_q,\G_q(H_q)])
\end{align*}
where $H_q(t)$ is the Hamiltonian of the geodesic curve $U_q(t)$, under the given $q$-metric. Plugging it into (\ref{Eq: dhdq}), we find
\begin{align*}
\dot K&=\frac{ d}{dq}\dot H_{q}(t)=-i\frac{ d}{dq}\Big(\F_q\big([H_q,\G_q(H_q)]\big)\Big)\\
&=-i\Big(\F_q\big([K,\G_q(H_q)]+\F_q\big([H_q,\G_q(K)]\big)\Big)-i
  \Big(\F_q\big([H_q,\P_\B(H_q)]\big)-1/q^2\P_\B\big([H_q,\G_q(H_q)]\big)\Big)  \\
& \equiv A(q,H_q,\F_q,\G_q,K) + M(q,H_q,\F_q,\G_q)
\end{align*}
where $A(q,H_q,\F_q,\G_q,K)$ is homogeneous and linear in $K$, while $M(q,H_q,\F_q,\G_q)$ is an inhomogeneous term which is not a function of $K$.  Thus we have derived a first-order ODE for $K(t)$, which is essentially a second order ODE for $J(t)$. We can express the solution $K(t)$ using a notation borrowed from dynamical systems. Since the above equation is a first-order linear equation for $K(t)$, we can define a linear operator $\K_t$ such that $\K_t(K(0))\equiv \K(K(0),t)=K(t)$ is the solution of the homogenous part of the above ODE. Then the solution of the entire equation including the inhomogeneous part $M(t)\equiv M(q,H_q,\F_q,\G_q)=-i\F_q^2\big([\P_\A(H_q),\P_\B(H_q)]\big)$ can be written as:
\begin{align*}
K(t)=\K_t(K(0))+\K_t\Big(\int_0^{t}ds\, \K_s^{-1}\big(M(s)\big)\Big)
\end{align*}
Next we substitute the expression of $K(t)$ into the integration form of $\dot J(t)=U_q^\dag(t)K(t)U_q(t)$:
\begin{align*}
\int_0^{T}dt\,U_q^\dag(t)K(t)U_q(t)=J(T)-J(0)=0
\end{align*}
and we get: 
\begin{align*}
&\int_0^{T}dt\,U_q^\dag(t)\K_t(K(0))U_q(t)
=-\int_0^{T}dt\,U_q^\dag(t)\K_t\Big( \int_0^{t}ds\, \K_s^{-1}\big(M(s)\big) \Big)U_q(t)
\end{align*}
where we have used the condition $J(T)=J(0)=0$. Let us define a linear operator $\J_T$, acting on $K(0)$:
\begin{align*}
\J_{T}\big(K(0)\big)=\int_0^{T}dt\,U_q^\dag(t)K(t)U_q(t)
\end{align*}
If $\J_T$ has an inverse, then we can express $K(0)$ as:
\begin{align*}
K(0)=-\J_T^{-1}\Big[\int_0^{T}dt\,U_q^\dag(t) \K_t \Big( \int_0^{t}ds\, \K_s^{-1}\big(M(s)\big) \Big)U_q(t)\Big]
\end{align*}
Moreover, we can further simplify the expression of the above right hand side. For $q=1$, the geodesic solution becomes a constant, $H_{q=1}(t)=H_{q=1}(0)$, and $\F_{q=1}$ and $\K_t$ are the identity operator; for $q>1$, we have the identity: $U_q^\dag(t) \G_q(H_q(t))U_q(t) =\G_q(H_q(0))$. Finally, together with the fact that $\dot J(0)=K(0)=\frac{ dH_{q}(0)}{dq}$, we obtain Eq.(\ref{dhdq}) as: 
\begin{align}\label{dhdq2}
\frac{ dH_{q}(0)}{dq}=\D\big(U_q(t),H_q(t)\big)
\equiv \left\{
\begin{array}{l}
\J_T^{-1}\Big(  \int_0^{T}dt\,U_q^\dag(t) it[\P_\A(H_q),\P_\B(H_q)] U_q(t) \Big),\, q=1;\\
\Big[\J_T^{-1}\big(\G_q(H_q(0))\big)T-\G_q(H_q(0))\Big]/\big(q(q-1)\big),q>1.
\end{array}\right.
\end{align}


\subsection{Numerical examples}


As illustrated in the main paper, we consider a two-qubit model with the following Hamiltonian, bounded by energy $E$:
\begin{equation}\label{heisen}
H = \hbar \! \sum_{l,m}  \! \omega_m^{(l)} \sigma_m^{(l)} + \hbar \kappa \sum_{m} \sigma_m^{(1)}\otimes\sigma_m^{(2)}
\end{equation}
where $\sigma_{m}^{(l)}$, $m = x, y, z$, $l=1,2$ are the Pauli operators for the $l^{\ms{\textit{th}}}$ qubit. 
Define $\A=\Span\{\sigma_m^{(l)},\sigma_m^{(1)}\otimes\sigma_m^{(2)}\}$ and $\B=\M/\A$, with $\dim\A=7$ and $\dim\B=8$. Let $\{A_j,B_k\}$ be the orthonormal basis for $\M=\A\oplus \B$. From controllability results, it can be shown that $H(t)\in \A$ is sufficient to generate arbitrary unitary gate in $\SU(4)$. Instead of directly solving the brachistochrone equation, which is extremely difficult, we will first solve the corresponding geodesic equation. So we relax the condition $H\in\A$ and assume $H(t)$ can be chosen from the entire space. Under the $q$-metric, we look for the geodesic solution $H_q(t)=(\alpha_j^q(t),\beta_k^q(t))$ for sufficiently large $q$, which then provides a good approximated solution for the corresponding brachistochrone equation. Then solving the brachistochrone equation becomes efficient. 

\subsubsection{Example 1: $U_{\tg}$ as a random unitary gate.}

We randomly choose a generic $U_{\tg}\in \SU(N)$, which is in the following form: 

\begin{align}\label{Ud_r2}
U_{\tg} =\begin{pmatrix}
-0.1479 + 0.3562i  & 0.0477 - 0.1303i  & 0.0508 - 0.7344i & -0.1364 - 0.5210i\\
  -0.0857 + 0.3357i  &-0.4268 + 0.0635i  & 0.5410 + 0.1276i & -0.5788 + 0.2233i\\
  -0.7706 + 0.0735i  &-0.1654 + 0.4709i  &-0.3602 - 0.0343i  & 0.0397 + 0.1390i\\
   0.3442 - 0.1166i & -0.2479 + 0.6957i  & 0.0372 + 0.1300i & -0.0088 - 0.5515i
\end{pmatrix}
\end{align}

For a fixed total time $T=1$, we can find the geodesic solution for $q=1$, through solving $H_{q=1}^0 = -i\log(U_{\ms{tg}})$. We get a sequence of solutions $\{\bar H^{(m)}\}$ with their norms in a nondecreasing order. The first four solutions of $\{\bar H^{(m)}\}$ are listed in Table~\ref{tab:log}. 

Next, for each $\bar H^{(m)}$, we apply geodesic deformation technique to solving Eq.(\ref{dhdq2}) and find the geodesic solution $H_q(t)$ for $q>1$. For $\bar H^{(1)}$ and $\bar H^{(2)}$, we find that Eq.(\ref{dhdq2}) can only be integrated to a finite value $q$, and hence we have to abandon these two branches. 

For $\bar H^{(3)}$ and $\bar H^{(4)}$, we can integrate Eq.(\ref{dhdq2}) from $q=1$ to $q=100$. For example, for $H_{q=1}^0 =\bar H^{(3)}$:

\begin{align*}
\bar H^{(3)} =
\begin{pmatrix}
-0.2920    &        -0.1913 + 0.2159i & -0.7479 - 0.8110i&   0.5540 - 0.5108i\\
-0.1913 - 0.2159i&  -0.2216   &          0.5196 - 0.3685i & -1.4427 - 0.6584i\\
-0.7479 + 0.8110i&   0.5196 + 0.3685i & -1.1955     &        0.0063 + 0.2534i\\
0.5540 + 0.5108i&  -1.4427 + 0.6584i&   0.0063 - 0.2534i &  1.7091         
\end{pmatrix}
\end{align*}

We can find $H_{q=100}^{0}$ from $H_{q=1}^{0}$ by integrating Eq.(\ref{dhdq2}), with simulation results illustrated in Table~\ref{tab:Ud-r2}. The fidelity of a geodesic solution $H_{q}(t)$ is defined as the final gate fidelity under the Hamiltonian control $H(t)\equiv \P_\A(H_q(t))$. From $H_{q=100}^{0}=\big(\alpha_j^{q=100}(0),\beta_k^{q=100}(0)\big)$, we get an approximated solution $\big(\mu_j(0),\lambda_k(0)\big)=\big(\alpha_j^{q=100}(0),q\beta_k^{q=100}(0)\big)$ of Eq.(\ref{eqn:brachicomp}), with a final gate fidelity $0.9916$. This approximated solution can then be used to seed the shooting method to solve Eq.(\ref{eqn:brachicomp}), giving a final gate infidelity as small as we like. For example, the brachistochrone solution $H(t)$ as shown in Table~\ref{tab:Ud-r2} has an infidelity less than $10^{-10}$, and $||H(t)||=6.69$ for $T=1$. Hence, after rescaling we obtain an optimal protocol time $T=6.69/E$. 

Starting from $H^{(4)}$, we can find a brachistochrone solution with $||H(t)||=7.49$ for $T=1$, with the optimal protocol time $7.49/E$ after rescaling. On the other hand, using weighted-sum optimization we find a rough upper bound $T^\ast=6.8/E$ for global optimal time. Repeating the above procedure for other larger values of $m$, with $||\bar H^{(m)}||<6.8$, we find other brachistochrone solutions, but all with $T^{m}>6.8$, implying that 
$T=6.69/E$ is very likely to be the global minimum time.

\subsubsection{Example 2: $U_{\tg}$ as the CNOT gate.}

Let's consider the target gate to be the CNOT gate:
\begin{align*}
U_{\tg} =\frac{\sqrt{2}}{2}(1-i)
\begin{pmatrix}
1   &     0     &   0    &   0 \\         
0    &   1 &   0   &   0  \\          
0   &    0  &    0  &1\\  
0    &    0 &   1&   0      
\end{pmatrix}
\end{align*}
where the global phase is added such that $U_{\tg} \in \SU(2)$.We fix the total time $T=1$. At $q=1$, we have $H_{q=1}^0=i\log(U_{\ms{tg}})$, which gives $[\P_\A(H_{q=1}^0),\P_\B(H_{q=1}^0)]=0$. This is the special case we mentioned in the main paper, and in order to solve Eq.(\ref{dhdq2}) we need to find $H_q^0$ at a value $q>1$. For example, at $q=5$, shooting method is still efficient to give a sequence of solutions $\{H_{q=5}^{(m)}(0)\}$ with their norms in a nondecreasing order. For each $m$, we can apply the deformation technique in order to find the geodesic solution at large $q$. For $m=1$, starting from $H_{q=5}^{0}=H_{q=5}^{(1)}(0)$, we can find $H_{q=100}^{0}$ by integrating Eq.(\ref{dhdq2}), with simulation details illustrated in Table~\ref{table3}. $H_{q=100}^{0}$ corresponds to a final gate fidelity of $0.9978$, and the approximated brachistochrone solution gives fidelity of $0.9974$. From there, we can seed the shooting method to find the accurate brachistochrone solution, with an optimal protocol time $5.75/E$. On the other hand, from weighted-sum optimization, we get an upper bound of the global minimum time, $T^\ast=5.8/E$, which will help identify the global time-optimal solution from numerics through comparison. 

\begin{table*}
\begin{tabular}{|c|c|c|c|}
\hline
$m$ & $\bar H^{(m)}=(\alpha_j,\beta_k)$ & $||\bar H^{(m)}||$  & $\alpha$\\
\hline
$1$ &    (0.3274    0.4584    0.5397    1.1585    0.1866    0.5210    0.8587    0.2493
    -0.0365   -0.9354    0.3499   -1.4261    1.0149   -0.3798    1.0106)  &  2.8783 & -1\\
$2$&   (-0.7359   -0.2306   -1.3681   -0.9518    1.8213   -1.7732    0.0636   -1.0186
   0.1108   -0.0134   -1.0328    0.2993   -0.6518   -0.0979   -0.0463)  &    3.5328 & i\\
$3$ & ( 1.4181   -0.1850   -0.4693   -1.4876   -2.1906    1.4694   -0.5136    0.8793
    -0.1975    0.1424    0.0375    0.6948    0.1526    0.3121   -1.0264)  &       3.7671 & -i\\
$4$ &  (-1.2809   -0.4489    1.2508    0.5144   -0.4232   -0.3323   -1.3300   -0.1244
    0.1477    1.8407    0.6265    1.8334   -1.4321    0.5882   -0.9263)  &       4.0205  & 1\\
    $\vdots$ &$\vdots$ & $\vdots$ & $\vdots$\\
\hline
\end{tabular}
\caption{Different solutions $\bar H^{(m)}=(\alpha_j,\beta_k)$ for $H_{q=1}^0 = -i\log(U_{\ms{tg}})$ with $U_{\ms{tg}}$ in~(\ref{Ud_r2}). } \label{tab:log}
\end{table*}

\begin{table*}
\begin{tabular}{|c|c|c|}
\hline
$q$ &Geodesic solution: $H_q(0)=(\alpha_j^q(0),\beta_k^q(0))$, $j=1,\cdots,7$, $k=1,\cdots,8$& Fidelity\\
\hline
1&   (1.2200   -0.1238   -0.6603   -1.3985   -2.4579    1.6768   -0.7312    0.4938    0.4424
   0.2142    0.1968    0.4047   -0.3723   -0.0584   -0.6108)  &   0.7612\\
2&   (0.9034    0.0337   -0.6488   -1.3562   -2.6288    1.7150   -0.8652    0.3377    0.4857
    0.1816    0.2110    0.3417   -0.4431   -0.1047   -0.5384)  &     0.7818\\
3 & ( 0.5851    0.1617   -0.6198   -1.3278   -2.7493    1.7308   -0.9837    0.2508    0.4839
    0.1704    0.2044    0.3025   -0.4521   -0.1243   -0.5086)  &       0.7992\\
4 &  (0.2747    0.2575   -0.5888   -1.3071   -2.8308    1.7388   -1.0848    0.1920    0.4702
    0.1691    0.1929    0.2698   -0.4434   -0.1351   -0.4932 )  &       0.8146\\
5&   (   -0.0214    0.3244   -0.5592   -1.2916   -2.8822    1.7421   -1.1688    0.1478    0.4525
0.1719    0.1804    0.2403   -0.4290   -0.1405   -0.4839)  &   0.8282\\
$\vdots$ &  $\vdots$&  $\vdots$\\
39 & ( -2.9985   -0.0496    0.8486   -0.3773   -2.3631    0.4896   -2.4435   -0.0778    0.2137 
0.1445   -0.0258   -0.0327   -0.1844   -0.0185   -0.4464)  &    0.9266\\
40  & (-2.9972   -0.0388    0.8888   -0.3574   -2.3680    0.4507   -2.4622   -0.0787    0.2103
   0.1420   -0.0280   -0.0340   -0.1811   -0.0166   -0.4463)  &   0.9273\\
$\vdots$ &  $\vdots$&  $\vdots$\\
59 &  (-2.8716    1.2846   -0.6363   -0.8567   -3.2048    2.0340   -1.4730   -0.1193    0.0566
 0.0177   -0.0016    0.0213   -0.0884    0.0074   -0.3967)  &      0.9559\\
60   &  ( -2.8693    1.2953   -0.6781   -0.8527   -3.2025    2.0753   -1.4622   -0.1201    0.0526
    0.0151   -0.0002    0.0237   -0.0866    0.0068   -0.3932) &     0.9567\\
99 & (-3.5774    0.5188   -2.4764   -0.0207   -1.7913    3.8783   -1.1532   -0.0784    0.0019
 -0.0244    0.0471    0.0426   -0.0362    0.0193   -0.1488)    & 0.9920\\
100 & (-3.5776    0.4989   -2.4919   -0.0148   -1.7645    3.8928   -1.1452   -0.0774    0.0019
   -0.0246    0.0471    0.0421   -0.0356    0.0192   -0.1465) & 0.9922\\
\hline
 & Brachistochrone solution: $(H(0),\lambda_k^q(0))=(\mu_j^q(0),\lambda_k^q(0))$, $j=1,\cdots,7$, $k=1,\cdots,8$& Fidelity\\
\hline
approx.& ( -3.5776    0.4989   -2.4919   -0.0148   -1.7645    3.8928   -1.1452   -7.7391    0.1918
   -2.4552    4.7084    4.2111   -3.5625    1.9163  -14.6530) & 0.9916\\
exact &(-4.0194    0.1372   -2.8829    0.2481   -1.0109    4.2998   -0.8674   -6.7600    0.0926   -3.0355
 5.9394    3.5790   -2.7144    3.3526   -9.7607)&     1\\
\hline
\end{tabular}
\caption{For $U_{\ms{tg}}$ in (\ref{Ud_r2}), geodesic solutions $H_q(0)$, $q=1,\cdots,100$, are calculated by integrating Eq.(\ref{dhdq2}) from $H_{q=1}^{0}=\bar H^{(1)}$. The brachistochrone solution $H(t)$ is found using shooting method with the good approximated solution derived from $H_{q=100}(t)$. } \label{tab:Ud-r2}
\end{table*}

\begin{table*}
\begin{tabular}{|c|c|c|}
\hline
$q$ & Geodesic solution: $H_q(0)=(\alpha_j^q(0),\beta_k^q(0))$, $j=1,\cdots,7$, $k=1,\cdots,8$ & Fidelity\\
\hline
5&   (-0.2517    1.5660    0.3099   -0.1429   -0.1428   -0.3098    1.5660    0.0624   -1.5488 -0.0535   -0.0624    0.0302    0.0535    0.0004   -0.0008)  &   0.5110\\
6&   ( -0.5907    1.5092    0.9173   -0.3865   -0.3863   -0.9173    1.5092    0.1531   -1.4052 -0.2006   -0.1531    0.0418    0.2006   -0.0095    0.0163)  &     0.5725\\
7 & ( -0.6899    1.4049    1.2875   -0.4685   -0.4684   -1.2875    1.4049    0.1840   -1.2924 -0.2869   -0.1840    0.0368    0.2869   -0.0125    0.0215)  &       0.6106\\
8 &  (-0.6641    1.0984    1.8222   -0.4269   -0.4269   -1.8222    1.0984    0.2278   -1.1489 -0.3584   -0.2278    0.0371    0.3584   -0.0066    0.0115)  &       0.6531\\
9 &  (-0.1509    0.0516    2.7287    0.0356    0.0353   -2.7287    0.0514    0.3031   -0.7892 -0.3732   -0.3031    0.0381    0.3732    0.0172   -0.0297)  &     0.7752\\
$\vdots$ &  $\vdots$&  $\vdots$\\
29 & ( -0.1255   -1.6023    3.2449    0.8386    0.8387   -3.2449   -1.6022    0.1118   -0.1267 -0.1143   -0.1118    0.0653    0.1143    0.0385   -0.0666)  &    0.9848\\
30  & (-0.1626   -1.6094    3.2497    0.8419    0.8419   -3.2497   -1.6093    0.1083   -0.1222 -0.1106   -0.1083    0.0654    0.1106    0.0382   -0.0661)  &   0.9857\\
$\vdots$ &  $\vdots$&  $\vdots$\\
 99 &  (-1.7415   -1.5536    3.2860    0.7932    0.7933   -3.2860   -1.5535    0.0331   -0.0403 -0.0363   -0.0331    0.0465    0.0363    0.0223   -0.0386)  &     0.9977\\
100   &  (-1.7644   -1.5519    3.2827    0.7921    0.7921   -3.2827   -1.5519    0.0328   -0.0399   -0.0359   -0.0328    0.0464    0.0359    0.0222   -0.0384) &     0.9978\\
\hline
 & Brachistochrone solution: $(H(0),\lambda_k^q(0))=(\mu_j^q(0),\lambda_k^q(0))$, $j=1,\cdots,7$, $k=1,\cdots,8$& Fidelity\\
\hline
 approx.&(-1.7644   -1.5519    3.2827    0.7921    0.7921   -3.2827   -1.5519    3.2827 -3.9922   -3.5942   -3.2827    4.6402    3.5942    2.2177   -3.8412) &     0.9974\\
exact &(-3.5651   -1.2154    2.8937    0.5839    0.5839   -2.8937   -1.2154    2.8937   -4.8797 -3.9232   -2.8937    7.3429    3.9232    3.2361   -5.6051)&     1\\
\hline
\end{tabular}
\caption{First, for $U_{\ms{tg}}$ as the CNOT gate, geodesic solution $H_{q=5}(0)$ is calculated using shooting method. Then $H_q(0)$, $q=5,\cdots,100$, are calculated by integrating Eq.(\ref{dhdq2}). Finally from the brachistochrone-geodesic connection, an accurate brachistochrone solution $H(t)$ is found, with $||H(t)||=5.75$.} \label{table3}
\end{table*}

\end{document}